\documentstyle[aps,pre,epsfig]{revtex}

\def\physd#1#2#3{{ Physica D}, {\bf #1}, #2 (#3)}
\def\prl#1#2#3{{ Phys. Rev. Lett.}, {\bf #1}, #2 (#3)}
\def\pla#1#2#3{{ Phys. Lett. A}, {\bf #1}, #2 (#3)}
\def\pra#1#2#3{{ Phys. Rev. A}, {\bf #1}, #2 (#3)}
\def\pre#1#2#3{{ Phys. Rev. E}, {\bf #1}, #2 (#3)}
\def\jpa#1#2#3{{ J. Phys. A}, {\bf #1}, #2 (#3)}

\def\ijbc#1#2#3{{ Int. J. Bifurcation and Chaos} {\bf #1}, #2 (#3)}

\def\eg{e. g.}
\def\etl{$et ~al.$~}

\def\R{$R$}
\def\h{$h$}
\def\beqr{\begin{eqnarray}}
\def\beq{\begin{equation}}
\def\eqn{\end{equation}\noindent}
\def\eqnr{\end{eqnarray}\noindent}

\def\plaa#1{{ Phys. Lett. A}, (#1)}
\begin{document}

\draft
\title{Intermittency transitions to strange nonchaotic attractors in
a quasiperiodically  driven Duffing oscillator}
\author{A. Venkatesan and M. Lakshmanan}
\address{Center for Nonlinear Dynamics, Department of Physics,
Bharathidasan University, Tiruchirappalli - 620 024, India}
\author{A. Prasad and R. Ramaswamy}
\address{School of Physical Sciences, Jawaharlal Nehru University,
New Delhi - 110 067, India}
\date{\today}
\maketitle
\begin{abstract}
Different mechanisms for the creation of strange nonchaotic attractors
(SNAs) are studied in a two--frequency parametrically driven Duffing
oscillator.  We focus on intermittency transitions in particular, and
show that SNAs in this system are created through quasiperiodic
saddle-node bifurcations (Type-I intermittency) as well as through a
quasiperiodic subharmonic bifurcation (Type-III intermittency).  The
intermittent attractors are characterized via a number of Lyapunov
measures including the behavior of the largest nontrivial Lyapunov
exponent and its variance as well as through distributions of
finite--time Lyapunov exponents. These attractors are ubiquitous in
quasiperiodically driven systems; the regions of occurrence of various
SNAs are identified in a phase diagram of the Duffing system.
\end{abstract}

\section{Introduction}

Interest in the dynamics of quasiperiodically driven systems has grown
in recent years largely due to the existence of novel behavior such as
strange nonchaotic dynamics.  The initial work of Grebogi \etl
\cite{R1} showed that with quasiperiodic forcing, nonlinear systems
could have strange nonchaotic attractors (SNAs), namely attractors with
a fractal geometry, but with non-positive Lyapunov exponent.  Subsequent
studies have dealt with a number of important issues pertaining to
theoretical as well as experimental aspects \cite
{R2,R3,R4,R5,R6,R7a,R7b,R8,R9,R10,R11,R12,R13,R14,R15} of SNAs.

While the existence of SNAs is firmly established, a question that
remains interesting is the mechanism or bifurcations through which
these are created from regular or chaotic attractors. To date a number
of different scenarios have been identified: these  include torus
doubling to chaos via SNAs\cite{R3}, fractalization of torus\cite{R4},
the re-emergence of torus doubling sequence and the birth of SNAs
\cite{R5}, the occurrence of SNAs via blow out bifurcation\cite{R6}, the
appearance of SNAs through Type-I\ intermittent phenomenon\cite{R7a}, or
Type-III intermittency\cite{R8}, and so on
\cite{R9,R10,R11,R12,R13,R14,R15}.

Scenarios for the formation of SNAs often have parallels in scenarios
for the formation of chaotic attractors.  The most common route to SNAs
is the gradual ``fractalization'' of a torus \cite{R4} where an
amplitude or phase instability causes the collapse of the torus which
progressively gets more and more wrinkled as a parameter in the system
changes, eventually becoming a fractal attractor. This is also the
least well understood mechanism for the formation of SNAs since there
is no apparent bifurcation, unlike the crisis--like torus collision
mechanisms identified by Heagy and Hammel \cite{R3} and Feudel, Kurths
and Pikovsky \cite{R10}. In the former instance, a period--doubled
torus collides with its unstable parent, while in the latter, a stable
and unstable torus collide at a dense set of points, leading to SNAs.
The quasiperiodic analogue of a saddle--node bifurcation gives rise to
SNAs through the intermittent route \cite{R7a}, with the dynamics
exhibiting scaling behavior characteristic of Type--I intermittency
\cite{R22}. The quasiperiodic subharmonic bifurcation, on the other
hand, gives rise to Type--III intermittent SNAs \cite{R8}, when a torus
doubled attractor is interrupted by a subharmonic bifurcation,
resulting in the inhibition of torus doubling sequence.

Our prime concern in the present work is to understand how a typical
nonlinear system, namely the Duffing oscillator, responds to a
quasiperiodic forcing and exhibits different dynamical transitions
involving SNAs. In particular, intermittencies through which SNAs are
formed have been investigated, besides the standard mechanisms like
fractalization and torus collision.  Although some of these mechanisms
have already been identified in certain maps and continuous systems
\cite{R3,R4,R7a,R7b}, in order to generalize  such dynamical transitions in
real physical systems, we undertake our investigation on a damped, two
frequency parametrically driven double--well Duffing
oscillator\cite{R16}
\begin{equation}
\ddot{x}+h\dot{x}-[1+A(R\cos t+\cos \Omega t)]x+x^3=0
\label{eq1}
\eqn
which is a well-suited model \cite{HD} for  buckled beam oscillations.
The simplest experimental realization of the above equation, the
magnetoelastic ribbon has been extensively studied \cite{R16,R17}, and
is the first system where SNAs were observed \cite{R13} with $\Omega$
chosen to be an irrational number. The existence of several routes to
SNAs in Eq.~(\ref{eq1}) suggests that there may be experimental
realizations of different types of SNAs which are deserving of further
study; our analysis here is thus of some experimental relevance.

The parametrically driven Duffing oscillator, Eq.~(\ref{eq1}), is 
a rich dynamical system, possessing a variety of regular, strange 
nonchaotic and chaotic
attractors. We concentrate on the intermittent transitions to SNAs and
the mechanism by which they arise in a range in $(R-h)$ parameter
space.  (In addition to the intermittency routes mentioned above,
Yal\c{c}inkaya and Lai \cite{R6} have shown that on--off intermittency
can also be associated with SNAs created through a blowout bifurcation,
when a torus loses its transverse stability\cite{R6}. This bifurcation
does not occur in the present system in the range of parameters
studied.)

SNAs represent dynamics which is intermediate between regular and
chaotic motion, and therefore they need to be distinguished from
chaotic attractors and quasiperiodic (torus) attractors. There are
several quantitative criteria that can be used to determine the strange
nonchaotic nature \cite{R1,R2,R3,R4,R5,R6,R7a,R7b,R8,R9,R10,R11,R12}. The
most direct characterization of SNAs is through the largest Lyapunov
exponent (which should be nonpositive) and the existence of a
nontrivial fractal structure. We employ these criteria in the present
work in order to identify strange nonchaotic behaviour: Lyapunov
exponents are calculated in the usual manner \cite {standard}, and the correlation
dimension (calculated through the standard Grassberger---Procaccia
algorithm \cite{grassberger}) is used to determine whether or 
not the attractor is
fractal. In addition to the fact that Lyapunov exponents are negative
on SNAs, the variance of the Lyapunov exponents on SNAs is also large.
(Here, by variance we mean the fluctuations in the measured value of
finite--time Lyapunov exponents, as calculated from several 
realizations of the dynamics; {see the Appendix}). This quantity 
also shows characteristic changes across
transitions to SNAs (particularly when there are crises) \cite{R7b}.

A host of other properties have been used to characterise SNAs, such as
the scaling of spectral features or the ``phase sensitivity'' \cite{R10}.
SNAs are characterised by specific signatures in their frequency
spectrum wherein they admit a power law relation: $N(\sigma) \sim
\sigma ^{- \alpha}, 1< \alpha < 2$, where the spectral distribution
function $N(\sigma)$ is defined as the number of peaks in the fourier
amplitude spectrum larger than some value $\sigma$. Another measure
\cite{R10} is based on the presence of a complicated path between the
real and imaginary Fourier amplitudes which reflects the fractal
geometry of SNAs. 

Finer distinctions among SNAs formed via different mechanisms can be
made by use of various measures, \eg~ the nature of the variation of
the Lyapunov exponents and its variance near the transition values of
the control parameter \cite{R7b}, the nature of the bursting and
scaling laws in the case of intermittent types SNAs \cite{R7a}, the
statistical properties of finite-time Lyapunov exponents \cite{trichy}
and so on.

In the next section, the birth of SNAs associated with the mechanisms
mentioned above are discussed. In particular, the transition of a
two--frequency quasiperiodic attractor $\rightarrow $ torus doubled
orbit $\rightarrow $ SNA (through type-III intermittency) $\rightarrow
$ chaos and the transition from chaos $\rightarrow $ SNA (through
type-I intermittency) $\rightarrow $ torus, besides the standard ones,
are shown to be operative in Eq.~(\ref{eq1}).  In Section-III, a number
of Lyapunov measures such as the behaviour of the largest Lyapunov
exponent, its fluctuations, the distribution of finite-time Lyapunov
exponents are used to characterize the transitions from two-frequency
quasiperiodicity to SNAs.  Our results are summarized in Section IV.

\section{ The Parametrically driven Duffing Oscillator}
For our analysis, Eq.~(1) can be rewritten as
\beqr
\dot{x}&=&y,  \\
\dot{y}&=&-hy+[1+A(R\cos \phi +\cos \theta )]x-x^3,\label{eq2}  \\
\dot{\phi} &=&1,~~~~\dot{\theta} = \Omega.
\eqnr
Note that the three equilibrium points of the system for $A=0$
correspond to $-x+x^3=0$, so that there are two stable fixed points
$x_{\pm }^s=\pm 1$ and an unstable fixed point at $x^u=0$.

Fig.~1(a) is the overall phase diagram for the system with the
parameters $A=0.3, \Omega = {\sqrt{5}+1}/2,$ fixed, while
$R$ and $h$ are varied. The dynamical equations are integrated
numerically  using a fourth-order Runge-Kutta algorithm with adaptive
step size. The dynamical states and transitions among them are
characterised through the Lyapunov exponents  and its variance as well as 
a number of different measures.  The relevant details of the calculation
of Lyapunov exponents and their variance are given in the Appendix.
 
There are a number of different regions where periodic, chaotic and
strange nonchaotic attractors can be found: Figs.~1(b), 1(c) and 1(d)
are blow--ups of regions W1, W2 and W3 respectively in Fig.~1(a).  The
various features indicated in the phase diagram are summarized and the
dynamical transitions are elucidated in the following.

The general features of the phase diagram fall into a familiar pattern.
Compared to the Duffing oscillator driven by a single frequency (the
case of $R = 0$), there are new chaotic regions C1, C2 and C3, and
bordering the chaotic regions, one has regions where the attractors are
strange and nonchaotic. The different regions where quasiperiodic
attractors can be found are also separated here by regions of chaotic
attractors and SNAs. In Fig.~1, the region marked TL contains
quasiperiodic attractors that oscillate about all the three equilibrium
points, $x^u, x^s_{\pm}$, while TS denotes a small quasiperiodic orbit
which  oscillates around one of the stable fixed points alone,
depending on the initial conditions. The strip denoted D contains
interesting dynamical states, both chaotic attractors and SNAs, between
which there are a number of transitions.

SNAs are found in a large number of regions, some of which are marked
GF1, GF2, GF3, HH1, HH2, S1 and S3 (based on the scenarios responsible
for their creation). It should also be pointed out that boundaries
separating different dynamical states are very uneven in this phase
diagram, which should be considered representative and schematic.  In
order to illustrate the fine transitions that take place in certain
regions when parameter varies, we also present the Lyapunov spectrum as
a function of $h$ for fixed $R$ in Fig.~2.  Fuller details are
discussed below.

Periodic orbits of the forced system with $R =0$ become quasiperiodic
tori of the system with nonzero $R$. As $R$ is further increased, these
tori typically bifurcate via period doubling. Upon further increase of
the parameters, there can be further bifurcations or other
transformations of the torus attractors to SNAs.

We first discuss the intermittency routes to SNA that can be observed
in this system.

\subsection{Type~~III intermittency}

In some regions in $(R-h)$ space the torus doubling sequence is tamed
due to subharmonic bifurcations which lead to the creation of SNAs
\cite {R8}.  We find that a growth of the subharmonic amplitude begins
together with a decrease in the size of the fundamental amplitude; such
behaviour is characteristic of the so--called Type~III intermittency
\cite {R22,R23}. The transition from a period--doubled torus to
intermittent SNA takes place in the region marked S3 in Fig.~1. From
Figs.~1(a) and 1(c), it can be observed  that initially the large
quasiperiodic orbit (TL) undergoes a torus doubling bifurcation to the
torus attractor denoted 2TL.  One would then expect the doubling
sequence to continue as in the usual period-doubling cascade, but
instead, here the doubling is interrupted by the formation of an
intermittent SNA which then finally settles into the chaotic attractor
C2 as $h$ is increased.

To understand the mechanism of the interruption of the doubling cascade
consider a specific parameter value of $R = 0.47$ while $h$ is varied
[see Figs.~2(a,b)].  For $h=0.08$, the attractor is a two-frequency
torus, but beyond this, further period doubling of doubled torus does
not take place.  Instead, a new dynamical behaviour, namely
intermittent phenomenon starts appearing at $h=h_c$=0.088689. This
transition is clearly illustrated in Figs.~3 where we  note  in the (x,
$\phi$ mod $4\pi$)\footnote{The various bifurcations and different
routes to SNAs can be easily identified by displaying $\phi$ modulo
$4\pi$ instead of modulo $2\pi$ \cite{HD}.} plane that the amplitude of
one of the components increases while the amplitude of the other
component decreases when a transition from doubled torus to
intermittent phenomenon takes place. To exemplify this transition
further in the Fourier spectrum, it has been identified that the
amplitude of the subharmonic component (W2/2) increases while the
amplitude of fundamental component (W2) {\it decreases} during this
transition (see Figs.~3).  This  suggests that the  birth of the
intermittent SNA is through a quasiperiodic analogue of the subharmonic
bifurcation. At this transition, the amplitude variation loses its
regularity (Figs.~3) and a burst appears in the regular phase. This
behaviour repeats as time increases as observed in the usual
intermittent scenario\cite{R22,R23}. Also, the duration of the laminar
phases (namely the quasiperiodic orbit) is random.  At the intermittent
transition the distinctive signature is an abrupt change in the
Lyapunov exponent as well as  in its variance as a function of $h$, as
shown in Fig.~4(a,b).  This type of SNA occurs in the region
$0.088689<h<0.088962$. On further increase of the value of $h$ from $h=
0.088963$, we find the emergence of a chaotic attractor (C2) shown in
Fig.~3(d), which though visibly similar to the SNA [see Fig.~3(c)] has a
positive Lyapunov exponent.

To confirm further that the SNA attractor,  Fig.~3(c),  is associated
with standard type-III intermittent dynamics, we plot the frequency of
laminar periods of duration $\tau$, namely N($\tau$) in
Fig.~4(c). This  obeys the scaling \cite{R23,R24}
\beq
N(\tau) \sim \left \{ {\exp (-4 \epsilon \tau) \over [1- \exp(-4
\epsilon \tau)]
 } \right \}  ^{0.5}.
\eqn
We find  $ \epsilon=0.009 \pm 0.0003 $ to give a best fit for
the present data.

\subsection{Type~~I intermittency}

On the right edge of the chaotic region C3, there is a transition from
a chaotic attractor to a SNA and then to a quasiperiodic torus, TS.
This transition proceeds via Type--I intermittency \cite{R7a} in the
region marked S1 in Fig.~1(a).

Consider the specific parameter value $R=0.35$ and vary $h$; for
$h=0.1907$, the attractor is chaotic [Fig.~5(a)], and as $h$ is
increased to $h=0.190833$, the chaotic attractor transforms to an SNA
shown in Fig.~5(b).  On increasing the value of $h$ further,
 an intermittent transition from the SNA  to a torus as shown in
Fig.~5(c) occurs at $h_c = 0.19088564\ldots$.  At this transition, the
abrupt changes in the Lyapunov exponent as well as its variance
[Fig.~6(a,b)] shows the characteristic signature of the
intermittent route 
to SNA as in the Type-III case discussed above. Here the SNA, hopping
between the two wells of the system, transforms to the small
quasiperiodic torus TS which oscillates in one of the
wells as in the case
of periodically driven Duffing oscillator \cite {R16}. Also, the plot
between the number of laminar periods N($\tau$) and the  period length
$\tau$ [shown in Fig.~6(c)] indicates  that after an initial  steep decay
there is a slight hump and then a  fall off to a small value of
N($\tau$). It also obeys the relation \cite {R23,R25}
\beq
 N(\tau) \sim { \epsilon \over 2c} \left \{
\tau+ \tan \left [ \arctan \left ({c \over \sqrt {\epsilon \over
u}}\right ) - \tau \sqrt{\epsilon u} \right ] - \arctan \left ({c \over
\sqrt { \epsilon \over u}} \right ) \tau - \tau^2 \sqrt{\epsilon \over
u}
\right \}, \eqn where $c$ is the maximum value of $x(t)$, $u=5.0$
and $\epsilon=0.0003 \pm +0.00002.$

\subsection{Other routes to SNAs}

In addition to the intermittency routes discussed above, the Duffing
system has the usual scenarios of torus collision as well as
fractalization.  The details are as follows.

\subsubsection {Torus collision}

Torus collisions---the route identified by Heagy and Hammel \cite{R3}--
are denoted HH in Fig.~1. In this scenario, a period$-$2$^n$ torus
attractor gets wrinkled and upon collision with its parent unstable
2$^{n-1}$ torus, a 2$^{n-1}$-band SNA is formed. Such a route has been
identified in two different regions (HH1 and HH2) of the present
system; the resulting SNAs have somewhat different morphologies.

To exemplify the nature of this transition, let us fix the parameter
\R~ at 0.3 while \h~ is varied. For \h=0.19, one finds the
presence of a small quasiperiodic attractor which is denoted as TS in
Fig.~1, but by \h = 0.185, the attractor undergoes a torus doubling
bifurcation and the corresponding region is denoted as 2TS in Fig.~1.
When \h~ is decreased further, the two strands of the doubled attractor
begin to wrinkle at \h $\approx$ 0.184 [Fig.~7(a)], and lose continuity
when \h $\approx$ 0.181, resulting in a fractal attractor as shown in
Fig.~7(b).  The strange nonchaotic [the Lyapunov exponent is negative
as seen in Fig.~8(a)] attractor is thus born at the collision of a
stable period doubled torus 2TS and its unstable parent TS.
Furthermore, this is a one band SNA, as is clearly depicted in Figs.~7,
in which prior to the collision, the 
dynamics in the (x, $\phi$ mod $2\pi$) plane corresponds to the two
branches [Fig.~7a(i)] while it corresponds to a simple curve in the (x,
$\phi$ mod 4$\pi$) plane [Fig.~7a(ii)]. However after the collision the
dynamics now goes over all of the attractor in both cases of (x, $\phi$
mod $2\pi$) [Fig.~7b(i)] and (x, $\phi$ mod $4\pi$) plane
[Fig.~7b(ii)].  

The Lyapunov exponent also shows distinct change in its behaviour for
this type of transition.  Fig.~8(a) is a plot of the maximal Lyapunov
exponent as a function of \h~ for a fixed value of \R~ =0.3. In the
neighborhood of $h_c$, $\Lambda$ varies smoothly in the torus region
but in the SNA phase the variation is irregular and the crossover
between these two behaviours is abrupt.  It is also possible to
identify the transition point from the examination of the variance in
$\Lambda$ [which increases significantly on the SNA as shown in Fig.~8(b)].

Similar behaviour has also been observed in the region denoted
HH2, for $ 0.37 < R < 0.5$ and $0.11 < h < 0.16$, where the large
period-doubled torus (2TL) becomes wrinkled and then interacts with its
unstable parent torus leading to the creation of a one band SNA.

\subsubsection {Fractalization}

The process of fractalization \cite{R4}, whereby a torus continuously
deforms and wrinkles to form a SNA can be seen in regions marked GF1--3
in Fig.~1.  The qualitiative (geometric) structure of the attractors
remains more or less the same during the process, unlike the
intermittency routes or like the torus collision route discussed in the
previous subsections. In this route, a period$-2^n$ torus becomes
wrinkled and then the wrinkled attractor gradually loses its smoothness
and forms a $2^n-$band SNA.

This transition can take place both as a function of increasing or
decreasing $h$ in different regions of the parameter space as can be
seen in the phase diagram. Consider the case $R= 0.47$, with varying
$h$.  For $h=0.05$, the attractor is chaotic [the Lyapunov exponent
(see Fig.~2) is positive]. As $h$ is increased to $0.0558$, one obtains
the attractor shown in Fig.~9(a) which is morphologically similar to
the chaotic attractor, but as it has a negative Lyapunov exponent
(Fig.~2) it is in fact an SNA.  On increasing \h~ to $0.065$, the
attractor loses its fractal character, becoming a (wrinkled) large
quasiperiodic orbit [see Fig.~9(b)]. The Lyapunov exponent ($\Lambda$)
and its variance ($\sigma$) do not show any specific signature as in
the case of the intermittent SNA or HH SNA [see Fig.~10(a)].

There are a variety of transitions involving fractalization in the
phase diagram.  On increasing $h$ to $0.072$, along $R=0.47$ in the GF2
region, the wrinkled attractor again loses its continuity and
ultimately approaches a fractal attrcator [see Figs.~9(c,d)], getting
considerably more wrinkled with increasing $h$. Beyond $h = 0.0725$,
the attractor becomes a doubled large quasiperiodic orbit with period
2TL.  This transition is illustrated in Figs.~7(c,d) (see also Fig.~2)
in which the 2TL orbit has two bifurcated strands of length 2$\pi$ for
$x>0$ which are actually a single strand of length 4$\pi$. However if
one looks in the reverse direction of $h$, one can note that as the 2TL
orbit loses its smoothness and develops a fractal nature, the dynamics
in the (x, $\phi$ modulo $2\pi$) plane and (x, $\phi$ modulo $4\pi$)
plane are geometrically the same as in the case of fractalization in
the GF1 region described above. The behaviour of the Lyapunov exponent
and its variance lack specific signatures at the torus to SNA
transitions as demonstrated in Fig.~10(b).

Yet another transition to SNA occurs through gradual fractalization in
the region GF3 where a transition from small quasiperiodic attractor
(TS) to chaos occurs through torus doubling bifurcation [Figs.~1(a) and
1(d)]. Here the period doubled orbit 2TS gets fractalized, leading to
the formation of the 2-band SNAs when $h$ is decreased from higher
values [see Fig.~1(d)] and the phenomenon is similar to what has been
discussed above.

\subsection {Transition between different SNAs}

Previous subsections have enumerated the several ways through which
SNAs are created from torus attractors. These processes include
transitions such as ...nT $\leftrightarrow$ $n$ band SNA
$\leftrightarrow$ chaos, ...$2^n$T $\leftrightarrow$ $2^{n-1}$ band SNA
$\leftrightarrow$ chaos, etc. One might also observe from the $(R-h)$
phase diagram, Figs.~1 that there are regions where transitions from
one type of SNA to another type can occur. Particularly, one may note
that transitions occur between (GF2 and S3), (S3 and HH2), (S1 and
HH1), and (HH1 and GF3). On closer scrutiny, we find that there exists
a narrow range of chaotic motion between the GF2 and S3 as well as the
S3 and HH2 regions, while 2TS/TS attractors intervene in between S1 and
HH1. That is, the transition between S3 and HH2 regions corresponds to:
.. $n$ band SNA $\leftrightarrow$ $n$ band chaos $\leftrightarrow$ $n$
band SNAs ... and between the regions GF2 and S3 corresponds to: ...
$n$ band SNAs $\leftrightarrow$ $n$ band chaos $\leftrightarrow$ ${n
\over 2}$ band chaos $\leftrightarrow$ ${n \over 2}$ band SNAs.
However in the transition region between GF3 and HH1, see Fig.~1(d),
there is band--merging, namely a transition from $n$--band SNA to
$n/2$--band SNA.  This transition \cite{negi} is analogous to the
phenomenon of reverse bifurcation or band merging bifurcations
occurring in chaotic systems, though here the dynamics remains
nonchaotic and strange during the transition. Such phenomena have been
identified in the driven Henon and circle maps and also in the logistic
map \cite{R7b,R12,negi}. In the present case we find that as the
parameter passes through critical values, the torus doubled attractor
does not collide in the GF3 region when undergoing fractalization,
whereas it does collide in the HH1 region, thereby leading to a
transition from one type of SNA to another (See Fig.~11).

\section{Probability distributions of Finite--time Lyapunov exponents}

Owing to the underlying fractal structure, trajectories on a SNA can be
locally unstable: finite--time Lyapunov exponents can be positive
although the global (or asymptotic) Lyapunov exponent is nonpositive.

As described in the Appendix, we obtain the distribution of finite
time Lyapunov exponents, $P(t,\lambda)$ and the variance, through 
Eq.~(A3) or (A4). 
For most ``typical'' chaotic motion, a general argument suggests that
the density $P(t,\lambda)$ [Eq.~(\ref{prob})], should be normally 
distributed \cite{gbp,R21}.
On the other hand, for intermittent dynamics, for instance, the system
switches between periodic and chaotic states and shows long-range
temporal correlations.  The probablity distribution of these
finite-time LEs is asymmetric since it arises from a superposition of
two independent densities, each of which is {\it separately} a Gaussian
distribution, centered on a distinct value of the average Lyapunov
exponent \cite{pre}. On intermittent SNAs, one of the components is a
torus with negative or zero Lyapunov exponent, while the other
component is chaotic, with average Lyapunov exponent being positive.
Finite stretches of intermittent dynamics involve both kinds of motion,
and therefore the density of finite--time LEs for intermittency usually
results in a stretched exponential tail \cite{pre} and in a
markedly asymmetric distribution for $P(t,\lambda)$. 

Although in the infinite time limit all distributions will go to a 
$\delta-$ function (see Eq.~(\ref{limit})), for finite times, 
$P(t,\lambda)$ can be very different for SNAs created through
different mechanisms \cite{R7b,trichy}. In particular, when the
dynamics is intermittent \cite{pre}, the exponential tails 
in the distribution persist for long times.
Shown in Fig.~12(a) are
distributions $P(2048,\lambda)$ for the two intermittent transitions
discussed here. 
Both the distributions have an asymmetric tail which
extends well into the locally chaotic ({\it i.e.} Lyapunov exponent
$\Lambda > $0) region, even for such a long time interval.
This correlates with enhanced fluctuations in
the Lyapunov exponent on intermittent SNAs. On SNAs formed by the
fractalization or torus collision mechanisms, in contrast, similar
densities are essentially gaussian \cite{trichy}; these are shown
for comparison in Fig.~12(b).

In order to quantify the slow decay of the positive tail in the distribution 
for intermittent SNAs, we
define  the fraction of postive local Lyapunov exponents \cite{pre} as
\begin {equation}
F_+(t)=\int_0^\infty P(t,\lambda) d\lambda.
\end{equation}
Clearly, lim$_{t \rightarrow \infty}F_+(t)\rightarrow 0$.
Empirically, it has been
found \cite {trichy,pre} that on the intermittent SNA, this quantity
shows the large $t$ behaviour
\begin{equation}
F_{+}(t) \sim t^{-\beta};
\end{equation}
we find numerically that the exponents in both the
S1 and S3 regions are $\beta \approx 1$ \cite{expo}.
For the fractalized or Heagy-Hammel SNAs, the approach is exponentially
fast,
\begin{equation}
F_{+}(t) \sim \exp(-\gamma t).
\end{equation}

Note that in the long--time limit, all distributions will collapse
to the Gaussian, so that for very long times, $t \to \infty$, 
$F_+$ will decay exponentially on all SNAs.  The exponents $\beta$ 
and $\gamma$ depend strongly on the system parameters.

\section{Conclusion}

On SNAs, as a consequence of the fractal geometry, stable and unstable 
regions are interwoven in a complicated manner. Thus, although
trajectories with different initial conditions will eventually coincide
with each other \cite{synchro} since there are no positive Lyapunov
exponents, they do so in an intermittent fashion, unlike
the case of quasiperiodic attractors,
converging in the
locally stable regions and diverging in the locally unstable regions. 

In this paper we have described the transition from quasiperiodic
attractors to chaos through different types of SNAs in a protypical
example, namely the two-frequency parametrically driven Duffing
oscillator. We have demarcated the different regions in parameter space 
where these different dynamical states can be located. 
There are several mechanisms through which SNAs are formed
here, two of which appear to be quasiperiodic analogies of
intermittency transitions in unforced systems \cite{R22}.  In addition
to these, we also find evidence for the torus collision route to SNA as
well as fractalization.

To distinguish among the different mechanisms through which SNAs are
born, we examined the manner in which the maximal Lyapunov exponent and
its variance changes as a function of the parameters. In addition, we
have also examined the distribution of local Lyapunov exponents and
found that on different SNAs they have different characteristics. Our
analysis confirms that in the intermittent SNAs, the signature of the
transition is a discontinuous change in both the maximal Lyapunov
exponent and  in the variance. Further, the two different intermittency
routes are distinguished by their different scaling behaviours.
The chaotic component on both types of
intermittent SNAs is long lived, giving, as a consequence, a slowly
decaying positive tail in P(N, $\lambda$) and a resulting power-law
decay for F$_+$(N).

Since the driven Duffing
system can be experimentally realized in the driven magnetoelastic
ribbon the present work can help in identifying different conditions
under which SNAs can be found in a typical system. The study of
SNAs---from both theoretical and experimental points of view---is in
its initial stages and the present study may aid in the
realization of  different bifurcation routes to SNA in experimental
systems.

\section {Acknowledgment} This work forms the part of Department of
Science and Technology, Government of India research projects to
ML and to RR (SP-S2-E07-96). A. V.
wishes to acknowledge Council of Scientific and Industrial Research,
Government of India, for financial support.

\appendix
\section{Finite-time Lyapunov exponents}
The largest Lyapunov exponent $\Lambda$ which measures the rate of separation
of nearby trajectories can be computed via a standard algorithm \cite{standard}. 
This is an asymptotic quantity, and is a long
(in principle infinite) time average of the local rate of expansion in
phase space. The finite--time or local Lyapunov exponent  $\lambda_i(t)$ is defined
in an analogus manner \cite{pre,abar} except that this is computed over a finite time
interval, $t$. The subscript $i$ indexes the segment (that is, in effect, 
the initial conditions)
in which this exponent is evaluated. A given trajectory is thus divided into 
segments of length $t$, and in each of these, the local Lyapunov 
exponent $\lambda_i(t)$ is computed.

As the finite--time Lyapunov exponents depend on the initial conditions,
we consider the probability distributions $P(t,\lambda)$, defined as
\beq
P(t, \lambda) d\lambda \equiv \mbox{ Probability that}
~\lambda_i(t)
\mbox{~takes a value between} ~~\lambda \mbox{~and~} \lambda + d\lambda.
\label{prob}
\eqn 
This distribution is a stationary quantity, and 
is particularly useful in describing the structure and dynamics of
nonuniform attractors \cite{gbp,R21}. In the asymptotic limit $t
\to \infty$, this distribution will collapse to a $\delta$ function,
\beq
P(t,\lambda) \to \delta (\Lambda - \lambda).
\label{limit}
\eqn
The deviations from this limit for finite
times, and the asymptotics, namely the approach to the limit can be
very revealing of the underlying dynamics \cite{pre}.

From this distribution, one can calculate the variance
of the Lyapunov exponent $\Lambda$ as
\beq
\sigma = \int_{ - \infty}^{\infty} (\Lambda - \lambda)^2 P(t,\lambda)
d \lambda.
\eqn
Equivalently, one can obtain the variance as
\beq
\sigma=\frac {1}{M} \sum_{i=1}^{M} (\Lambda-\lambda_i(t))^2,
\eqn 
namely from a set of $M$ finite--time Lyapunov
exponents. In our numerical calculations in Sec. II, for instance, 
we take $t=50$ and $M \cong 10^5$.

\begin{figure}
\caption{Phase diagram for the parametrically driven Duffing
oscillator system, Eqs.~(2)-(4), in the
$(R-h)$
parameter space. Here 2TS and 2TL correspond to torus doubled
attractors of small and large  quasiperiodic orbits,
respectively. GF1, GF2 and GF3 correspond to the regions where the
process of gradual fractalization of torus occurs. HH1 and HH2
represent the regions where SNA is created through the
Heagy-Hammel route.
S1 and S3 denote regions where the SNA appears through Type-I  and
Type-III intermittencies respectively. C1, C2 and C3 correspond to
chaotic attractors. Regions W1, W2 and W3 in Fig.~1(a) are expanded
in  Figs.~1(b), 1(c) and 1(d), and the  inset in Fig.~1(b)
illustrates the intertwining of chaotic (C1) and quasiperiodic
orbits (TL); the two levels curves correspond to the specific
values of the maximal Lyapunov exponent:
$\Lambda$=0 ($...$) and $\Lambda$=-0.005 ($---$).}
\end{figure}

\begin{figure}
\caption{The behaviour of the Lyapunov exponent as a function of $h$
for (a) $R=0.47$, (b) $R=0.47$ and (c) $ R=0.4$.
Here the notation is the same as in Fig.~1
(Tr stands for transients).}
\end{figure}

\begin{figure}
\caption{Projection of the two-frequency attractors of Eqs.~(2)-(4) for
R=0.47, with
the Poincar\'e plot in the (x, $\phi$) plane (i) $\phi$ modulo
2$\pi$
and (ii) $\phi$ modulo 4$\pi$ (iii) power spectrum indicating
the transition from chaotic to quasiperiodic attractors via SNA through
type-III intermittent  mechanism:
(a) two frequency doubled quasiperiodic orbit  for $h = 0.075$; (b)
doubled torus for
$h=0.08$; (c) intermittent SNA for $h=0.088689$; (d) chaotic attractor
for $h
=0.088963$.}
\end{figure}

\begin{figure}
\caption{Transition from doubled torus to SNA through type-III
intermittent
mechanism in the region S3: (a) the behaviour of the
Lyapunov exponent ($\Lambda$); (b) the variance ($\sigma$); (c)
Number of laminar periods $N(\tau)$  of
duration $\tau$ in the case of transition through type-III
intermittency.}
\end{figure}

\begin{figure}
\caption{Projection of the two-frequency attractors of Eqs.~(2)-(4) for
R=0.35, with
the Poincar\'e plot   in the (x, $\phi$) plane (i) $\phi$ modulo 2$\pi$
and (ii) $\phi$ modulo 4$\pi$ indicating
the transition from quasiperiodic to chaotic attractors via SNA through
type-I intermittency mechanism:
(a) chaotic attractor for $h=0.1907$; (b) intermittent SNA for
$h=0.190833$;
(c) torus for $h=0.19088564$.}
\end{figure}

\begin{figure}
\caption{Transition from   torus to SNA through type-III
intermittent
mechanism in the region S3: (a) the behaviour of the
Lyapunov exponent ($\Lambda$); (b) the variance ($\sigma$); (c)
Number of laminar periods $N(\tau)$  of
duration $\tau$ in the case of transition through type-I
intermittency.}
\end{figure}

\begin{figure}
\caption{Projection of the two-frequency attractors of Eqs.~(2)-(4) for
$R=0.3$, with
the Poincar\'e plot   in the (x, $\phi$) plane (i) $\phi$ modulo 2$\pi$
and (ii) $\phi$ modulo 4$\pi$ indicating
the transition from chaotic to quasiperiodic attractors via SNA through
Heagy- Hammel (HH1) mechanism:
(a)  torus doubled  attractor for $h=0.184$ ;
(b) strange nonchaotic attractor for $h = 0.181$.}
\end{figure}

\begin{figure}
\caption{Transition from doubled torus to SNA through Heagy-Hammel
mechanism in the region HH1: (a) the behaviour of the
Lyapunov exponent ($\Lambda$); (b) the variance ($\sigma$).}
\end{figure}

\begin {figure}
\caption{Projection of the two-frequency attractors of Eqs.~(2)-(4) for
$R=0.47$, with
the Poincar\'e plot  in the (x, $\phi$) plane (i) $\phi$ modulo 2$\pi$
and (ii) $\phi$ modulo 4$\pi$ indicating
the transition from chaotic to quasiperiodic attractors via SNA through
fractalization (GF1 and GF2).
(a) strange nonchaotic attractor for $h = 0.0558$ (GF1); (b) wrinkled
attractor for $h=0.065$; (c) strange nonchaotic
attractor for $h = 0.072$ (GF2); (d) torus doubled attractor for $h=0.0729$.}
\end{figure}

\begin{figure}
\caption{Transition from   torus to SNA through Gradual fractalization
mechanism indicating (i) the behaviour of the
Lyapunov exponent ($\Lambda$); (ii) the variance ($\sigma$):
(a) in the region GF1; (b) in the region GF2.}
\end{figure}

\begin{figure}
\caption{
Transition from gradual fractalization (GF3 region) type of SNA
to Heagy-Hammel (HH1 region) type of SNA for $h=0.1675$: (a) SNA (GF3
region) for $R=0.1955$;(b) SNA (GF3
region) for $R=0.197$; (c) SNA (HH1 region) for $R=0.1974$;
(d) SNA (HH1 region) for $R=0.203$.}
\end{figure}

\begin{figure}
\caption{Distribution of finite--time Lyapunov exponents on
SNAs created through (a) Type--III (dashed line) and Type--I intermittency
(solid line), with $t= 2048$, and (b) 
 torus collision (the Heagy--Hammel route) (dotted line) and 
 fractalization (solid line). 
The parameter values are those of
Figs.~3(c), 5(b), 7(b) and 11(a) respectively.}
\end{figure}
%\end{document}
\newpage
\centerline{\epsfig {figure=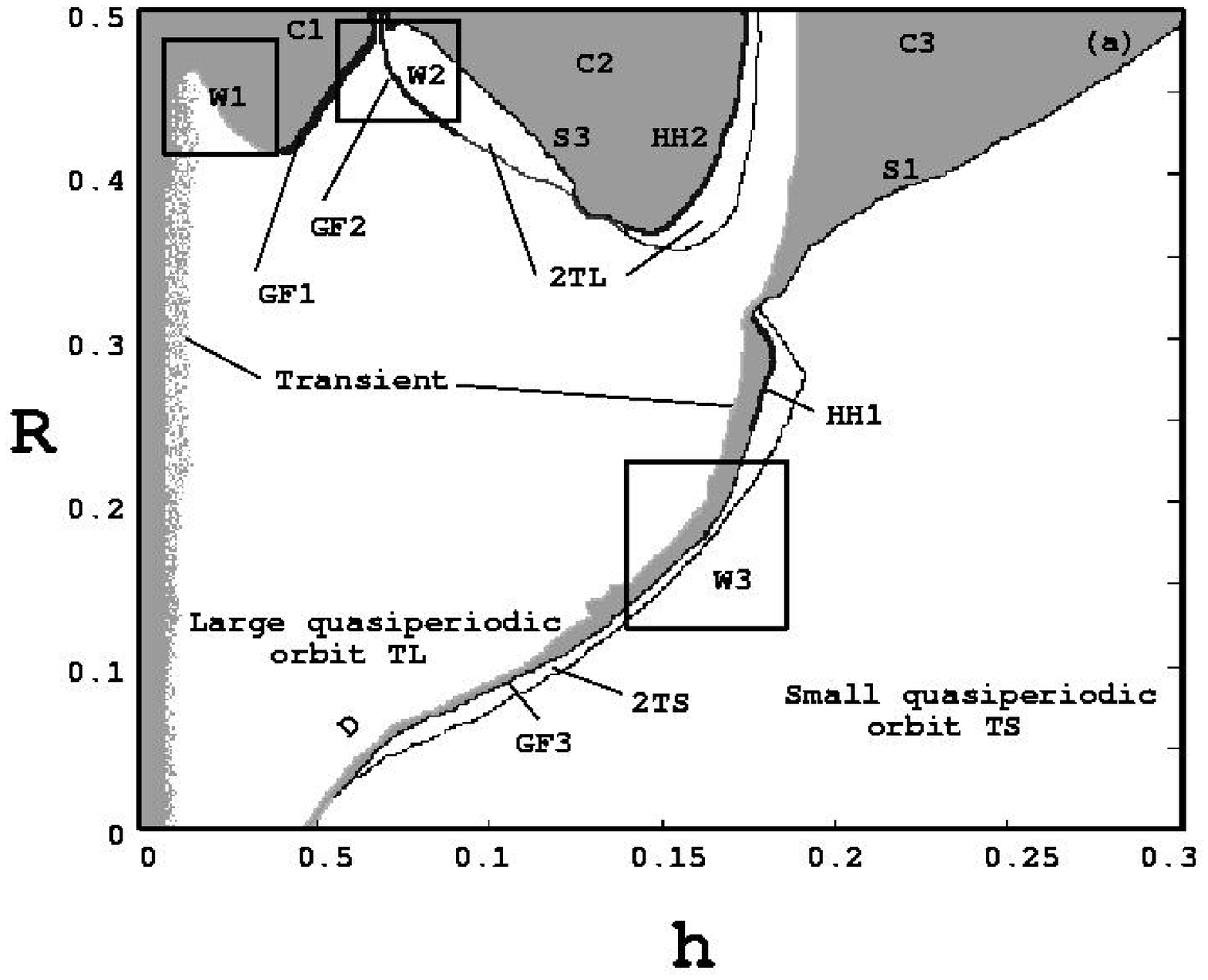,width=.99\linewidth}}
\centerline{FIG. 1a (A. Venkatesan et al.) }
\centerline{\epsfig{figure= 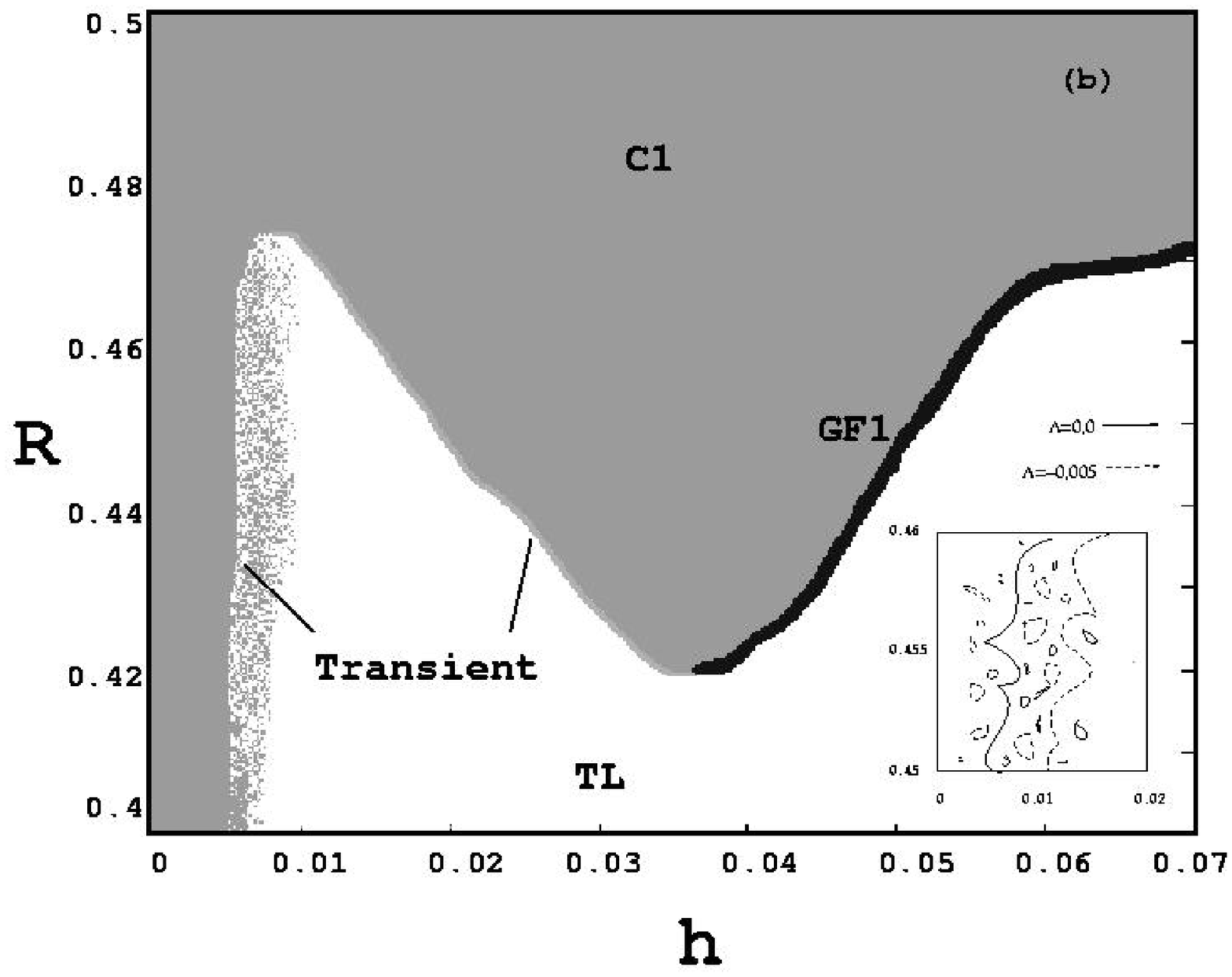,width=.99\linewidth}}
\centerline{FIG. 1b (A. Venkatesan et al.) }
\centerline{\epsfig {figure= 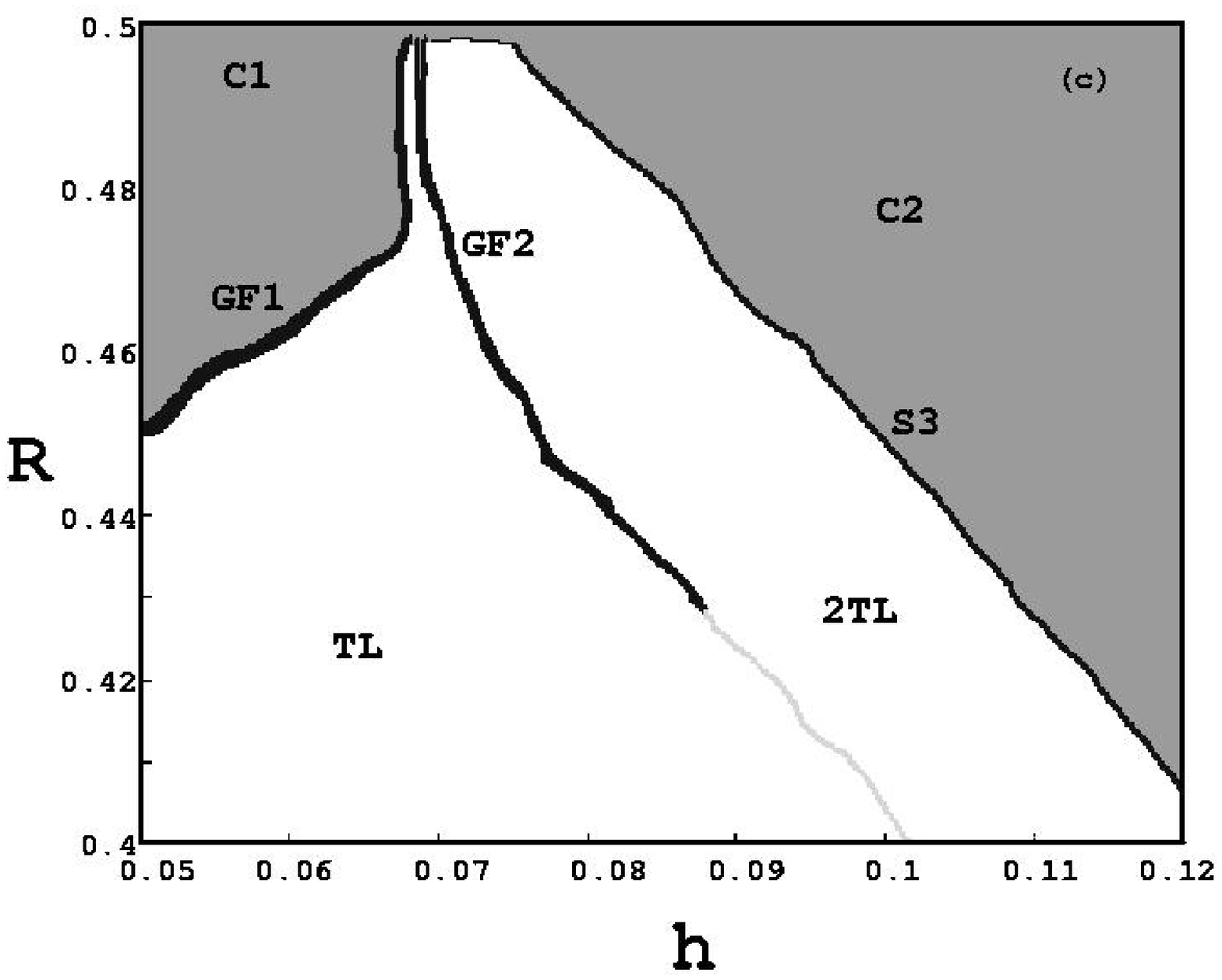,width=.99\linewidth}}
\centerline{FIG. 1c (A. Venkatesan et al.) }
\centerline{\epsfig{figure= 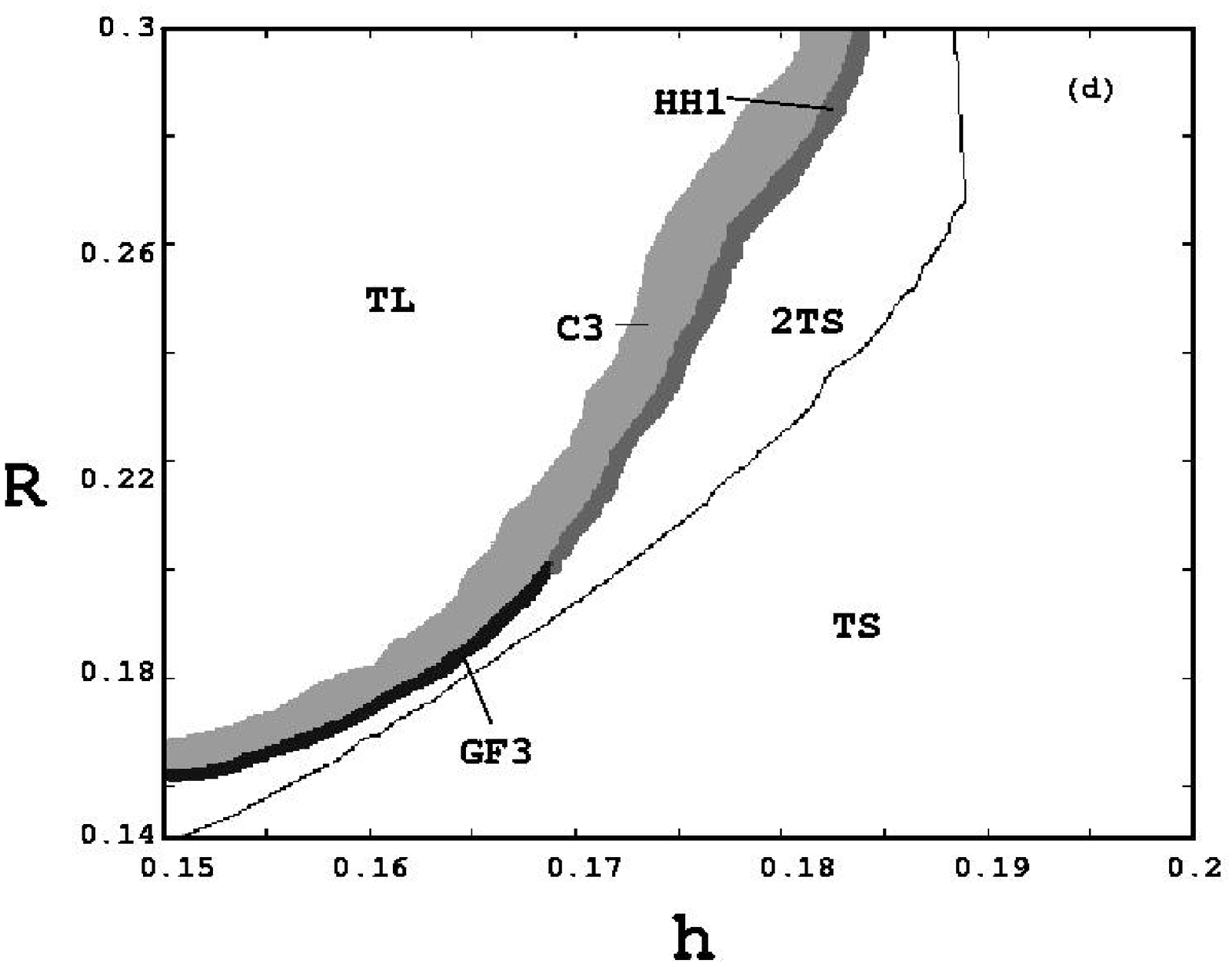,width=.99\linewidth}}
\centerline{FIG. 1d (A. Venkatesan et al.) }
\newpage
\centerline{\epsfig {figure=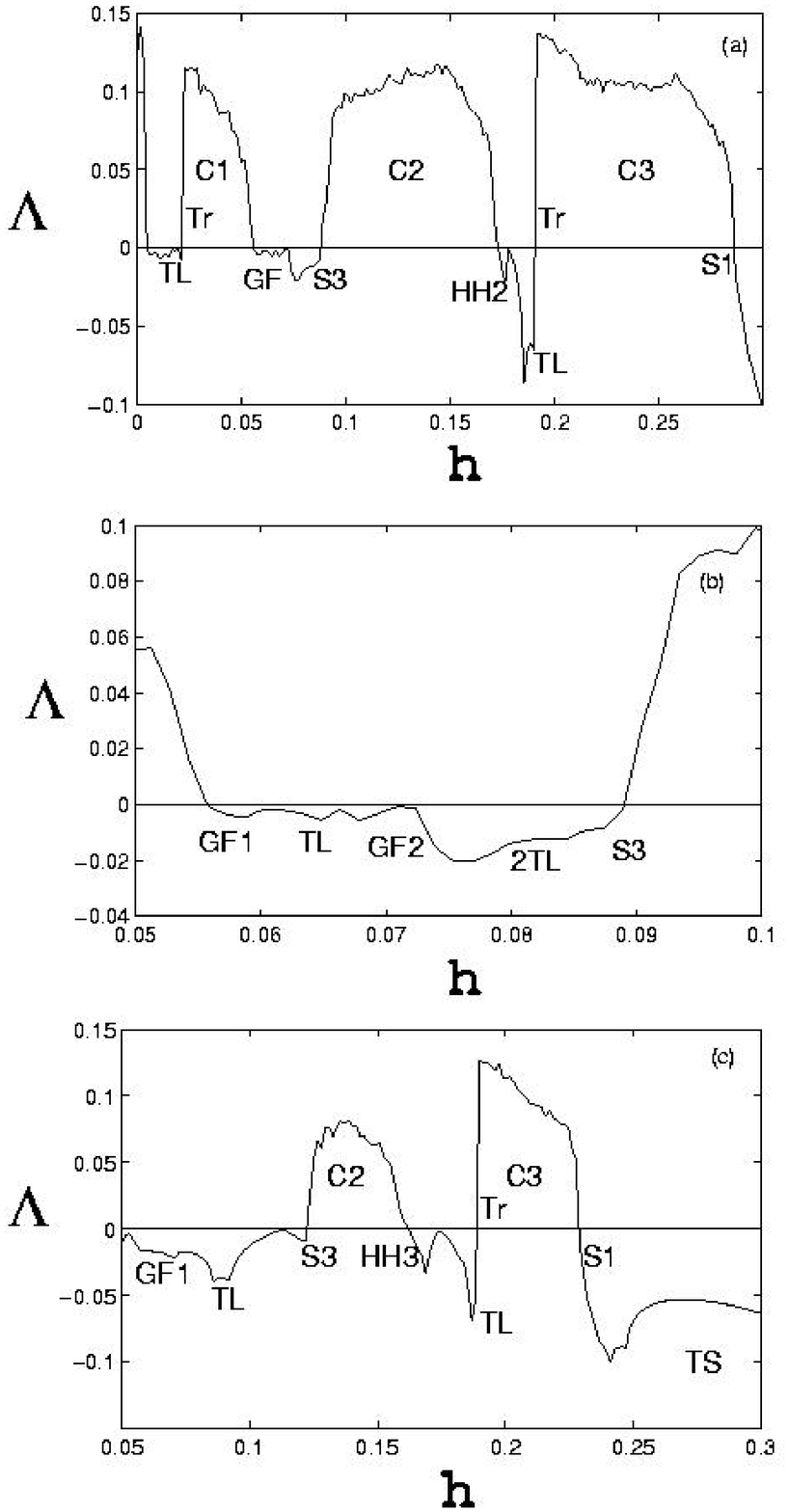, width=.6\linewidth}}

\centerline{FIG. 2 (A. Venkatesan et al.) }
\newpage
\centerline{\epsfig {figure=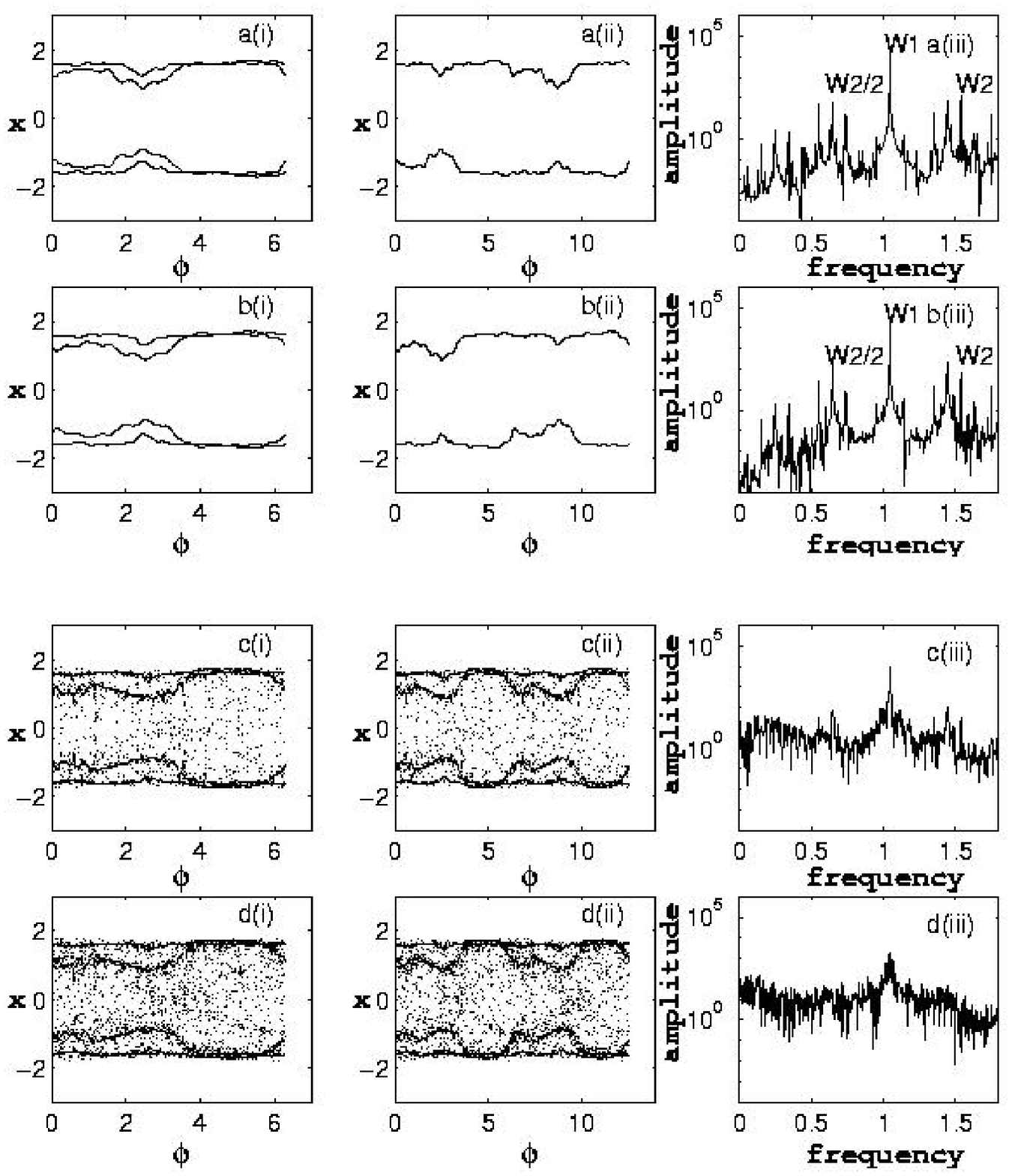, width=.9\linewidth}}
\centerline{FIG. 3 (A. Venkatesan et al.) }
\newpage
\centerline{\epsfig {figure=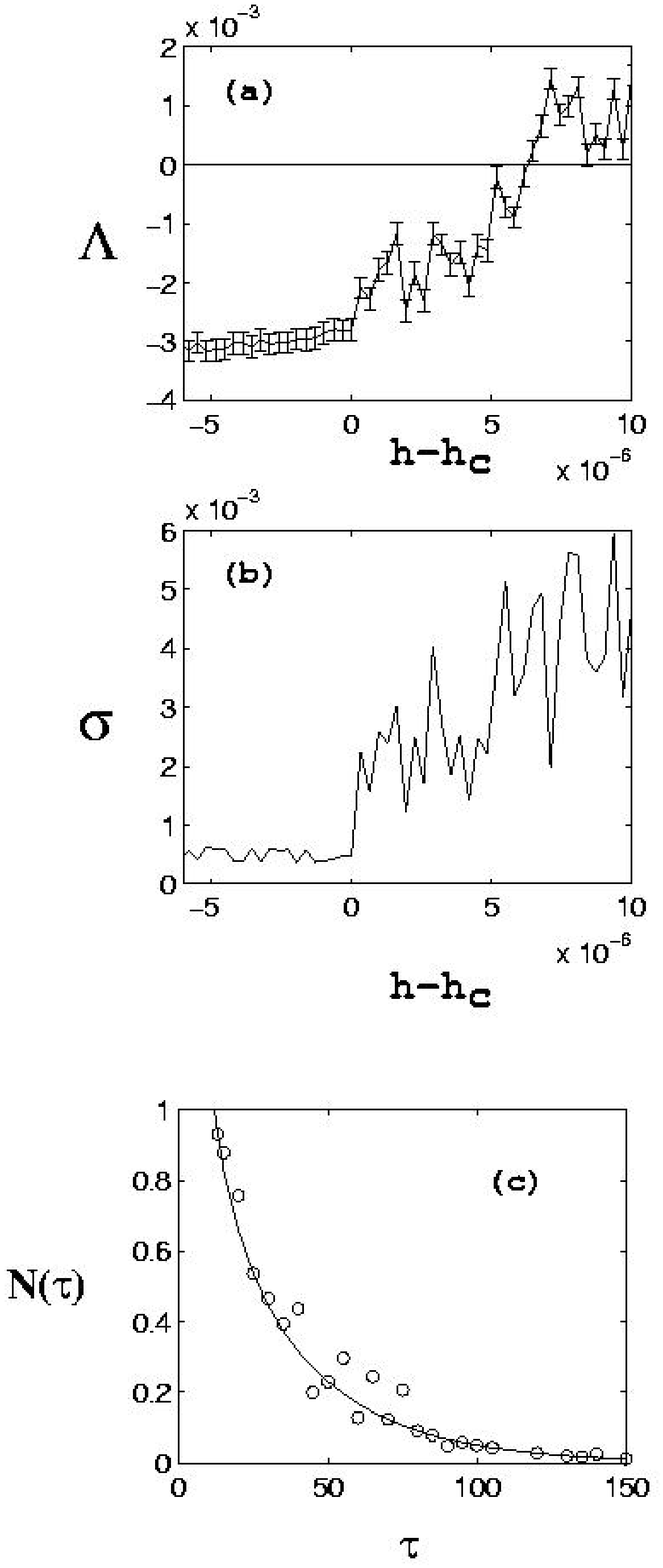, width=.5\linewidth}}
\centerline{FIG. 4 (A. Venkatesan et al.) }
\newpage
\centerline{\epsfig {figure= 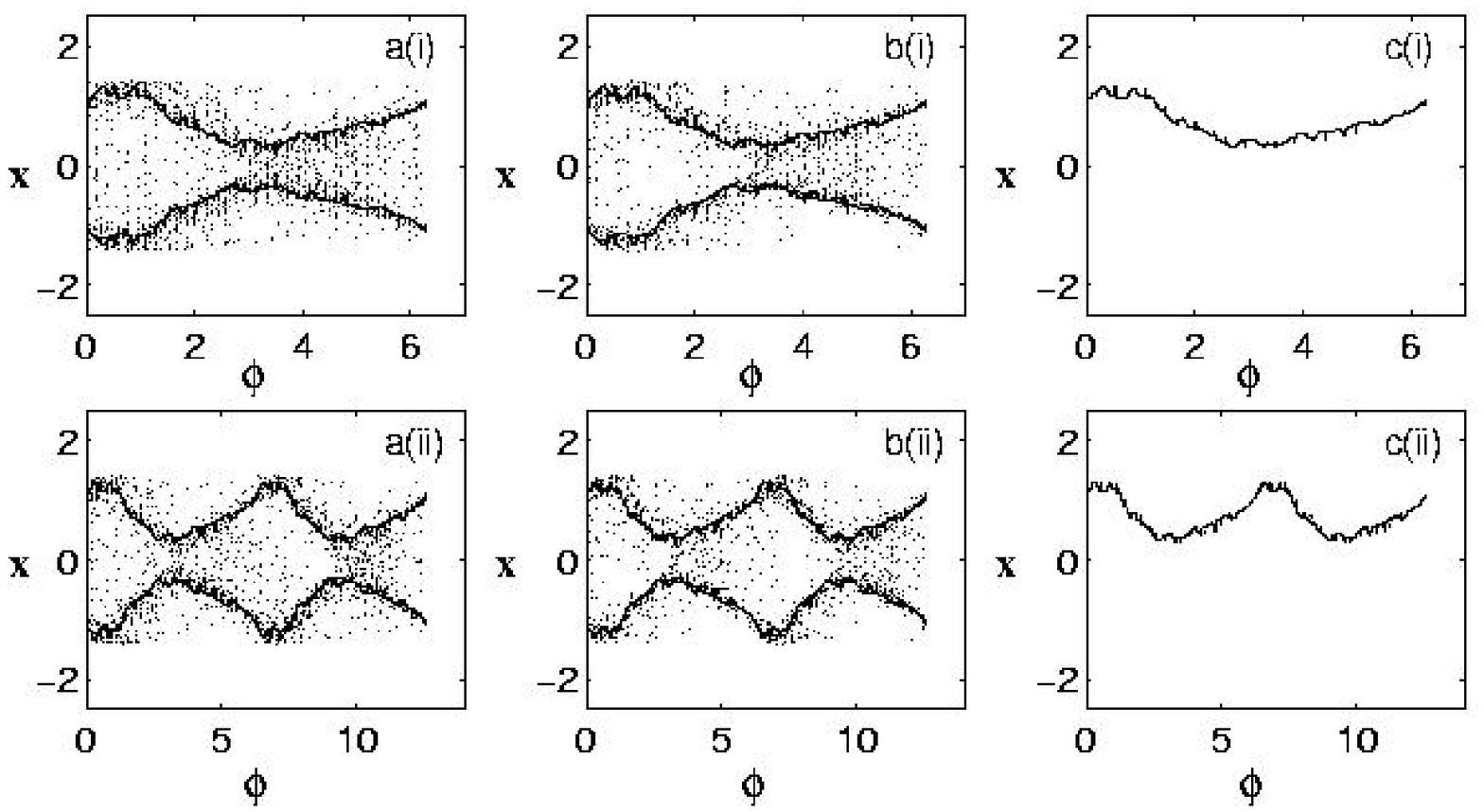, width=.9\linewidth}}
\centerline{FIG. 5 (A. Venkatesan et al.) }
\newpage
\centerline{\epsfig {figure= 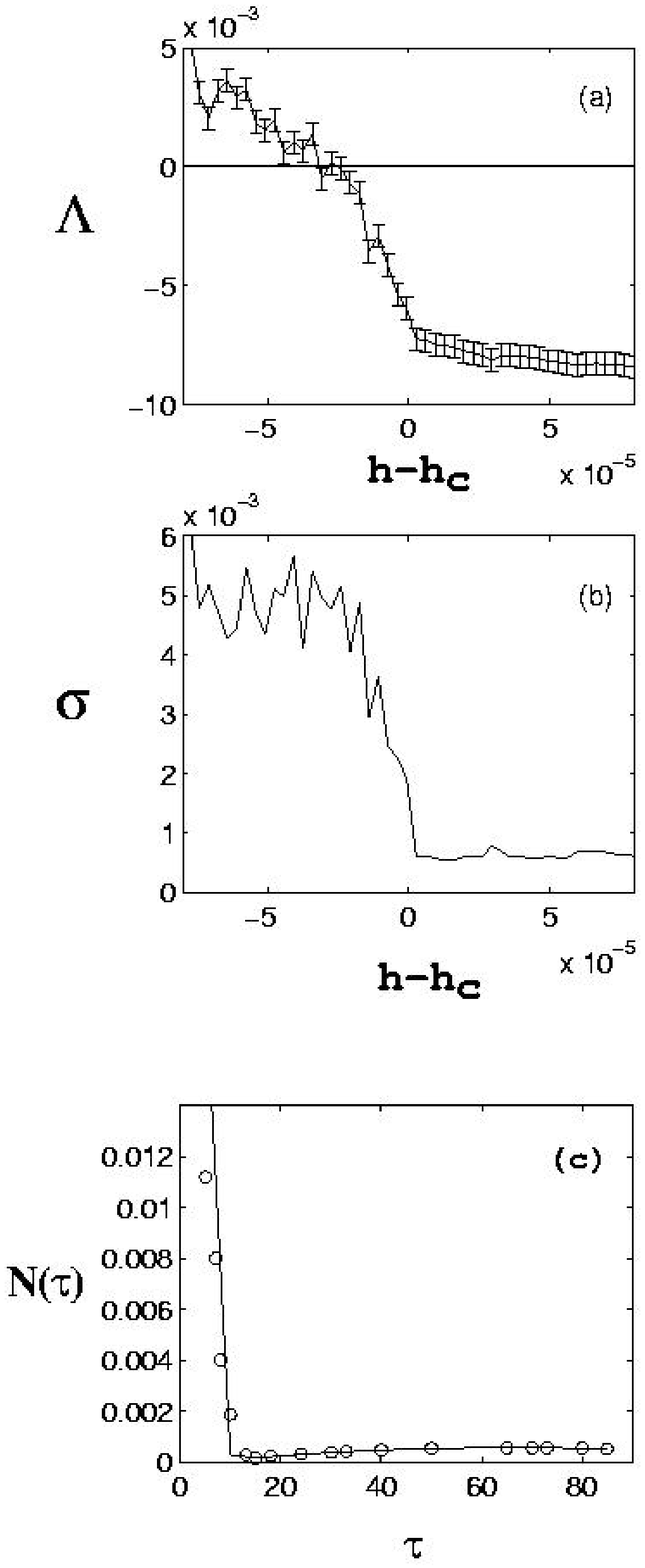, width=.5\linewidth}}
\centerline{FIG. 6 (A. Venkatesan et al.) }
\newpage
\centerline{\epsfig {figure= 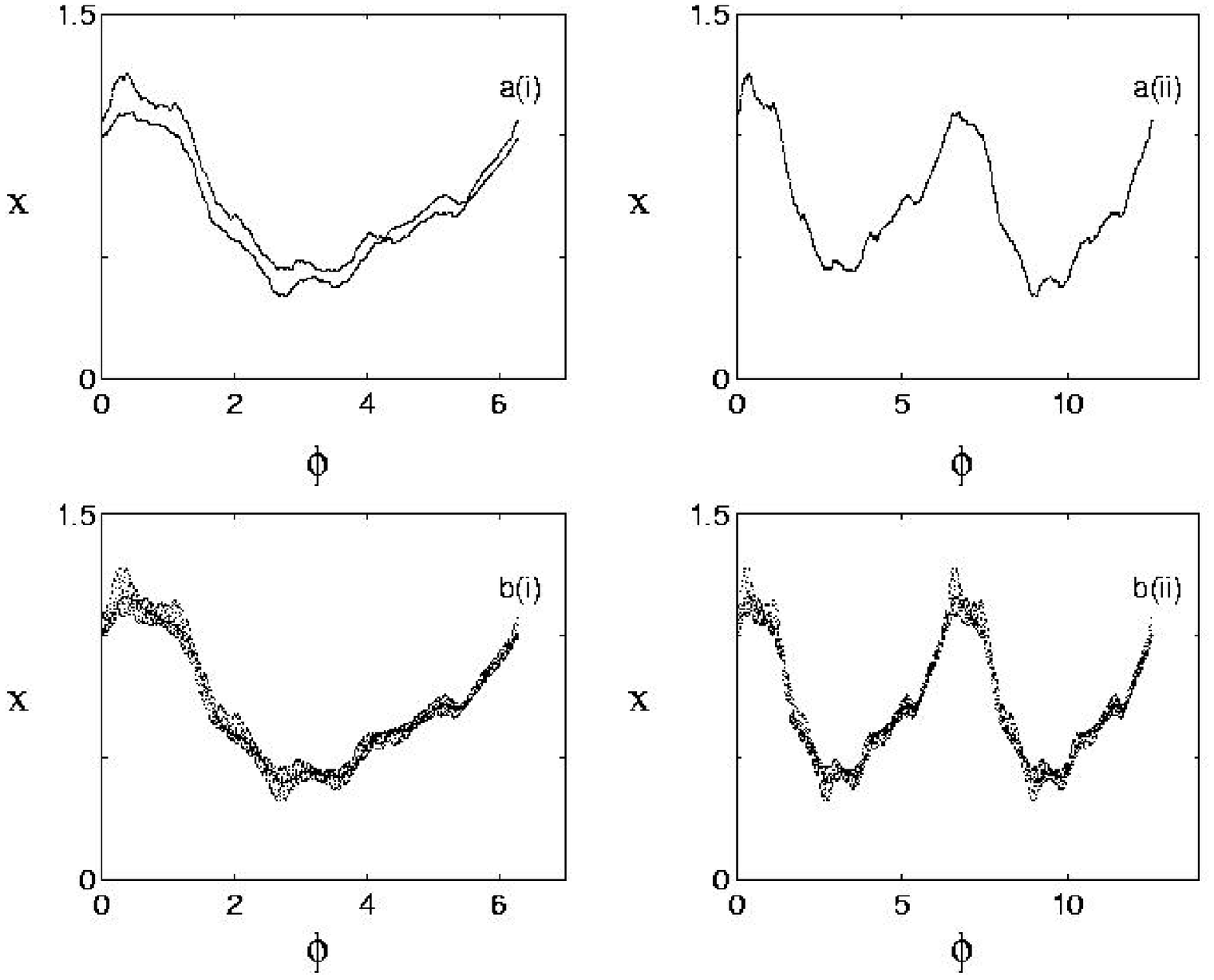, width=.8\linewidth}}
\centerline{FIG. 7 (A. Venkatesan et al.) }
\newpage
\centerline{\epsfig {figure= 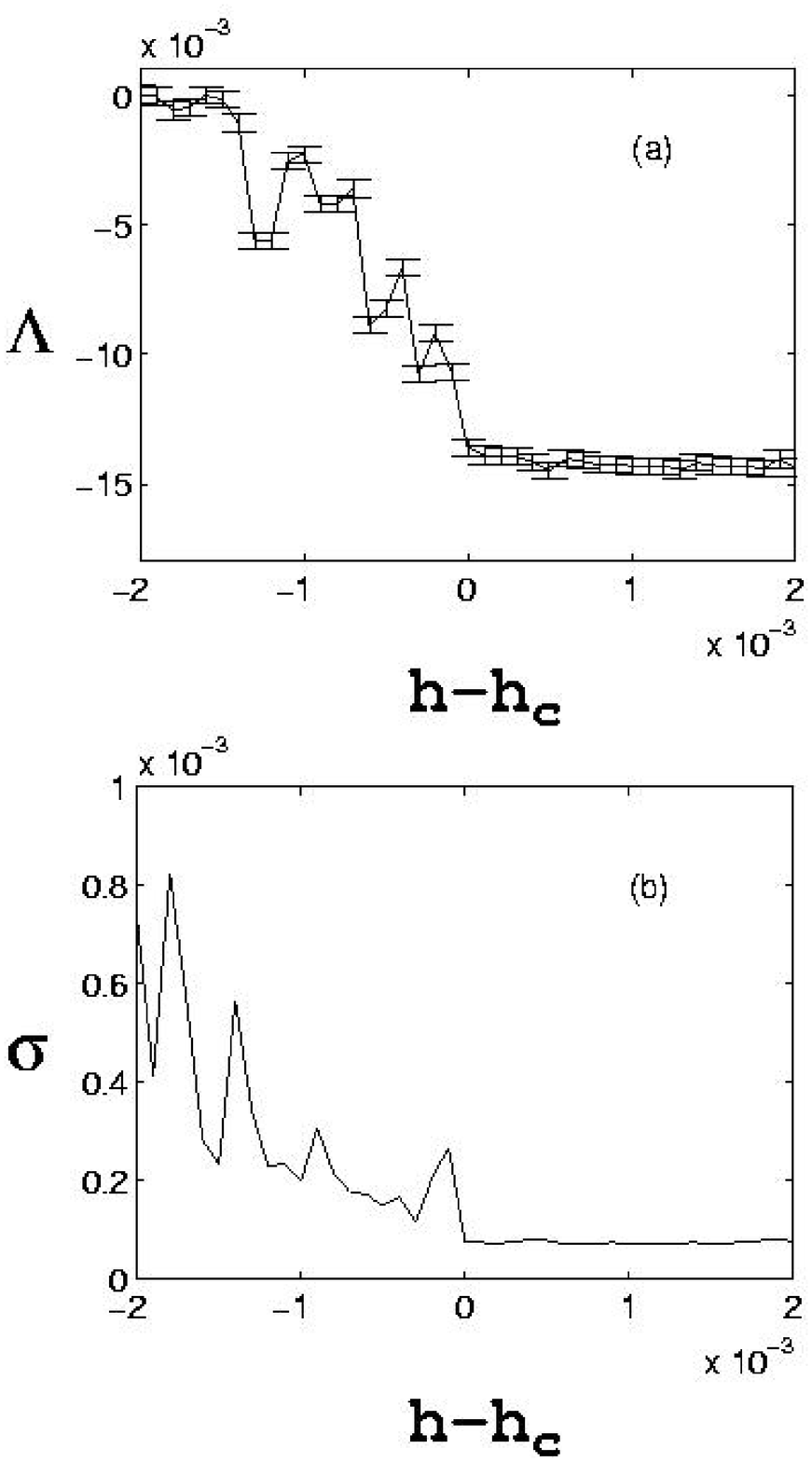, width=.7\linewidth}}
\centerline{FIG. 8 (A. Venkatesan et al.) }
\newpage
\centerline{\epsfig {figure= 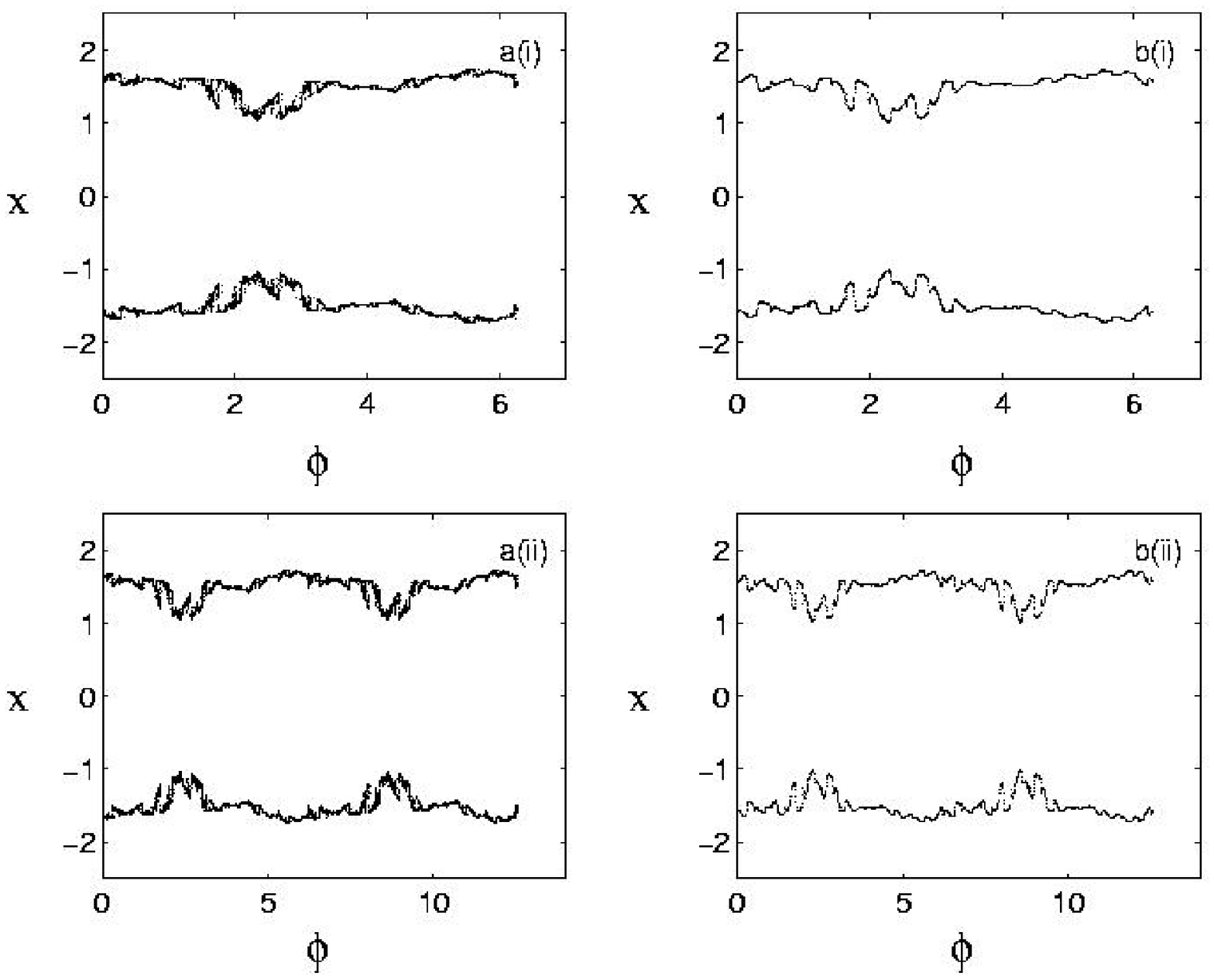, width=.8\linewidth}}
\centerline{\epsfig {figure= 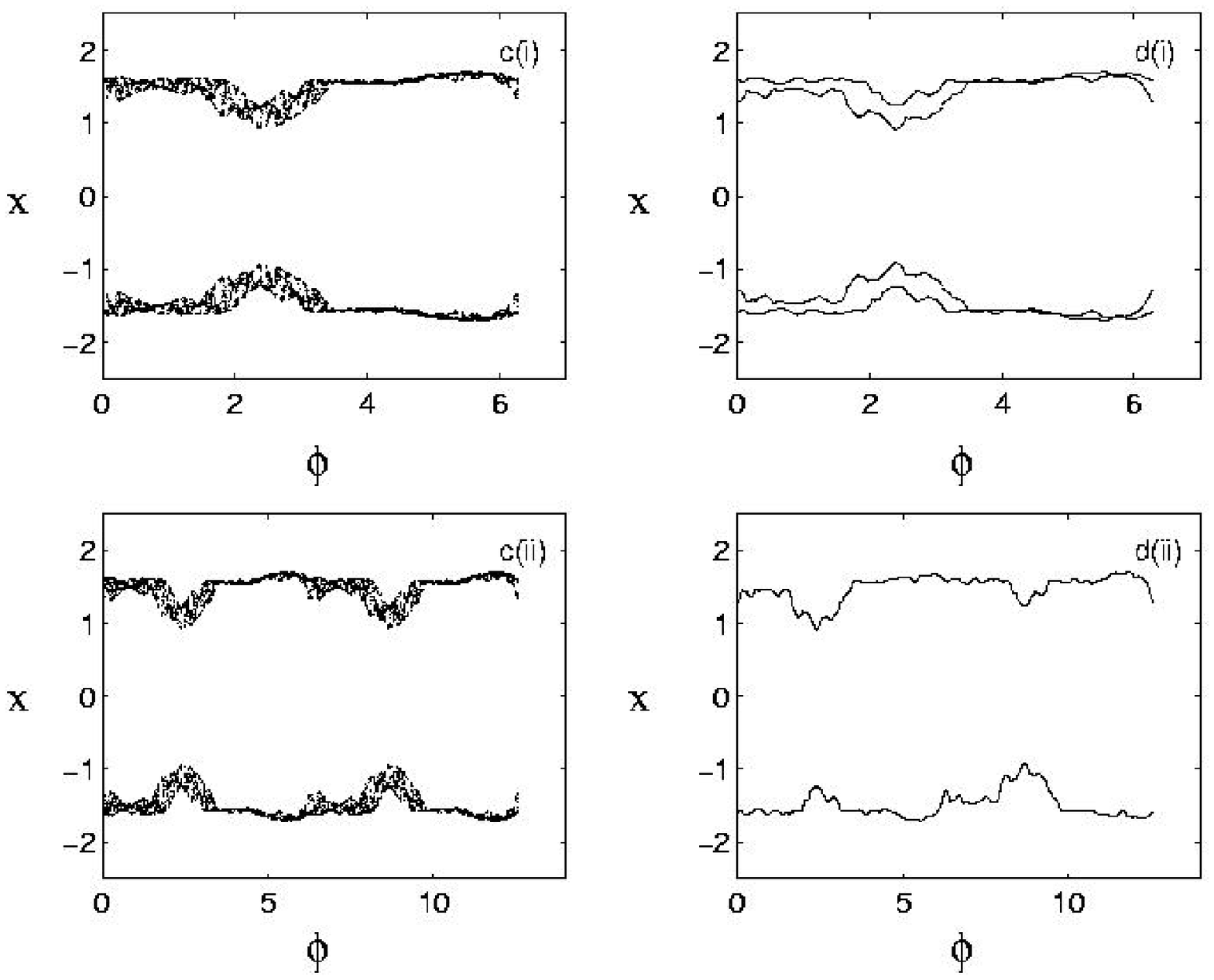, width=.8\linewidth}}
\centerline{FIG. 9 (A. Venkatesan et al.) }
\newpage
\centerline{\epsfig {figure= 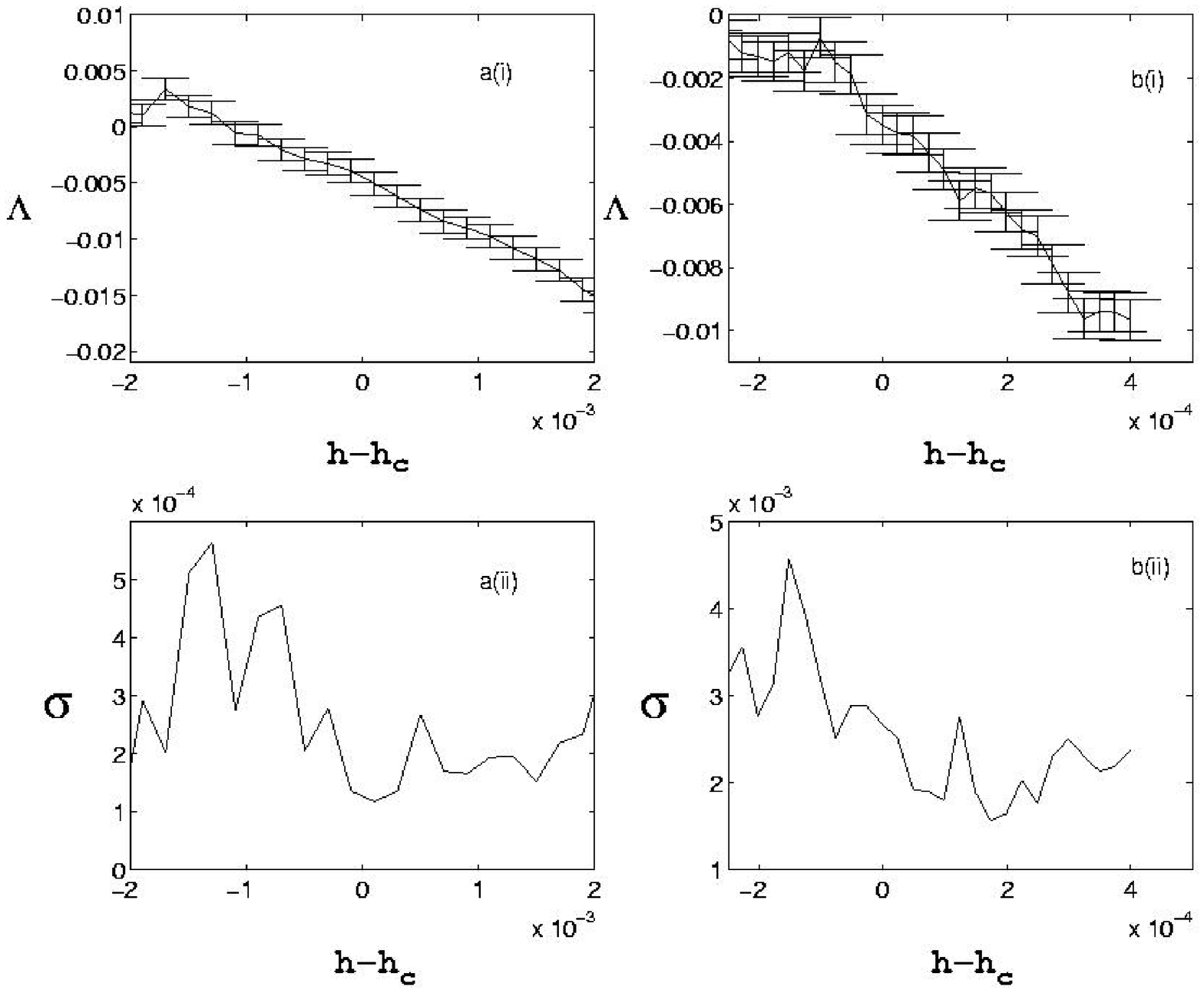, width=.99\linewidth}}
\centerline{FIG. 10 (A. Venkatesan et al.) }
\newpage
\centerline{\epsfig {figure=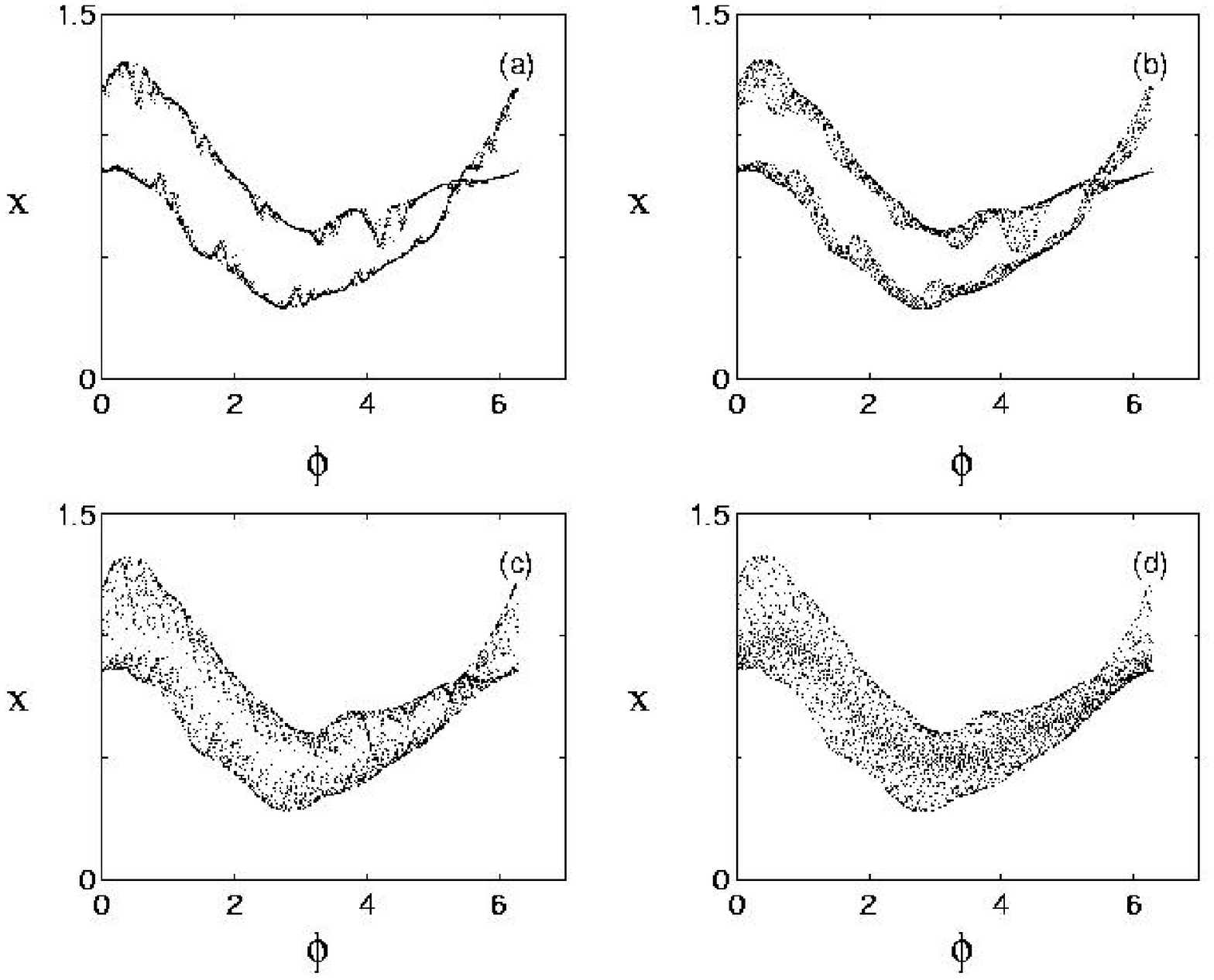, width=.8\linewidth}}
\centerline{FIG. 11 (A. Venkatesan et al.) }
\newpage
\centerline{\epsfig {figure=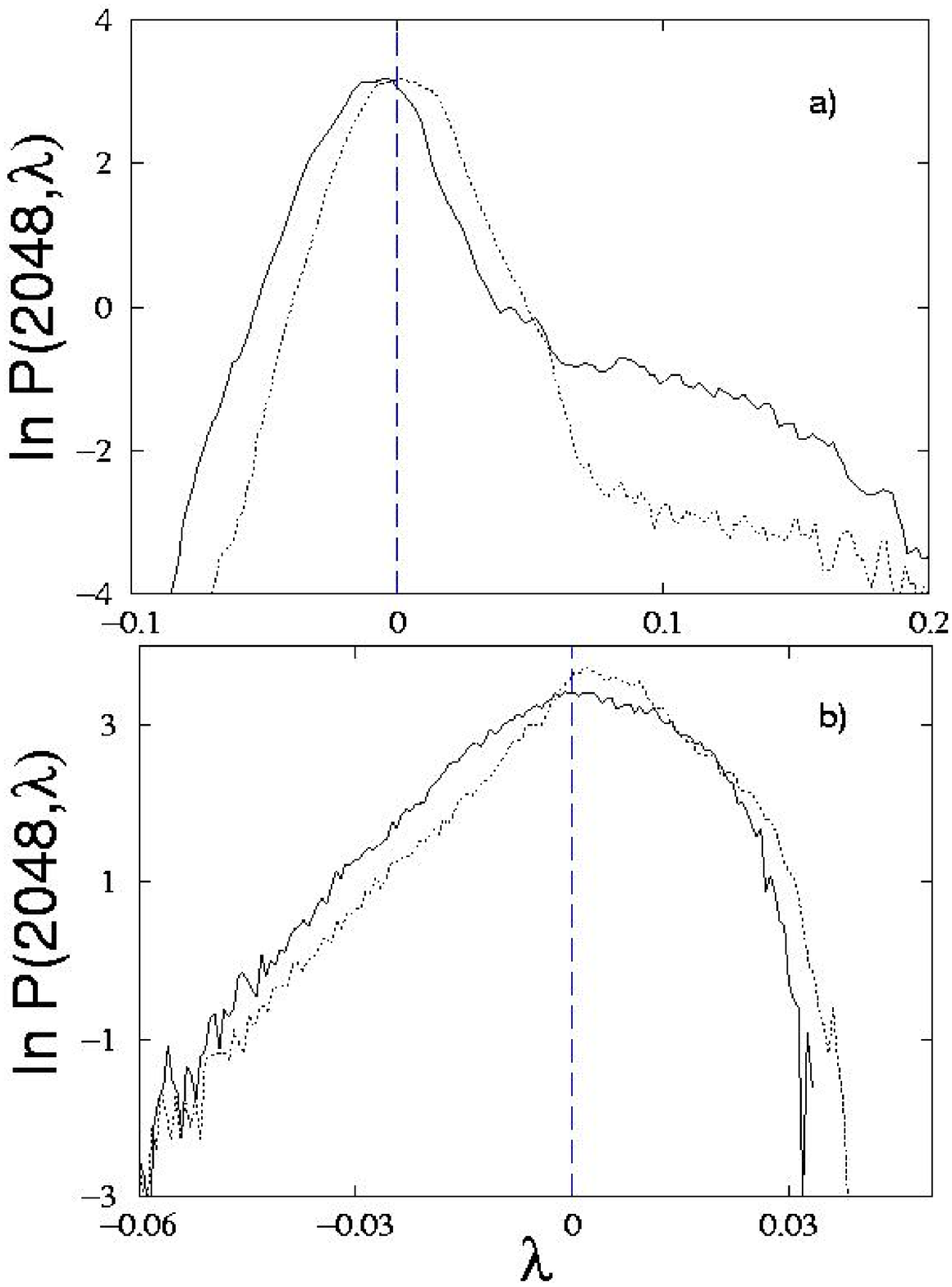, width=.8\linewidth}}
\centerline{FIG. 12 (A. Venkatesan et al.) }

\begin{references}
\bibitem{expo} At $R=0.45$ and $h=0.100736782$ (in the S3 region) the
exponent $\beta$ is $1.14$ and in the S1 region, at $R=0.35$ and
$h=0.190865$, the exponent is $0.89$.
\end{references}
\end{document}